\documentclass{article}

\usepackage{amssymb,amsmath,amsfonts,bm,tabulary,xcolor,graphicx,xspace,colortbl,rotating}

\begin{document}

\title{Material Line Fluctuations Slaved to Bulk Correlations in Two-Dimensional Turbulence}
\author{Theo Odijk\\
Lorentz Insitute for Theoretical Physics\\
University of Leiden\\
Niels Bohrweg 2\\
2333 CA Leiden\\
The Netherlands\\
E-mail: odijktcf@online.nl}
\date{\ }
\maketitle

\begin{abstract}
An analogy is pointed out between a polymer chain fluctuating in a two-dimensional nematic background and a freely floating material line buffeted by a two-dimensional turbulent fluid in the inertial (Kraichnan) regime. Under certain conditions, the back-reaction of the line on the turbulent flow may be neglected. The fractal exponent related to the size-contour relation of the material line is connected to a ``nematic'' correlation function in the bulk.
\end{abstract}

Theories of turbulence generally focus on the properties of correlation functions rather than make an attempt to solve the Navier-Stokes equation as such \cite{1,2,3}. If we restrict
ourselves to two dimensions [2D], it is the inertial regimes that are important at asymptotically high Reynolds numbers, at scales larger than the Kraichnan dissipation length $\lambda _{k}$\ \cite{4,5}. These regimes have been studied experimentally in soap films flowing under gravity in set-ups that allow for continuous operations \cite{6}. An interesting correlation function was measured a decade ago \cite{7}. A thin column of water was injected in a turbulent soap film: this could be viewed as a material line being deformed by the 2D turbulence. Amarouchene and Kellay succeeded in measuring the configurational statistics of the evolving fluctuating line \cite{7}. Here, I attempt to connect the line correlation function to that of the bulk turbulence albeit under conditions without symmetry breaking.

The problem is reminiscent of a polymer chain being deformed by a nematic matrix in two dimensions \cite{8,9}. Let us recall the argumentation used to connect the polymer correlation function with the underlying nematic correlations in the limit of strong coupling. The latter are expressed in terms of the director $\overrightarrow{n} (\overrightarrow{r}) \equiv \exp [i \, \theta (\overrightarrow{r})]$ where the angle $\theta (\overrightarrow{r})$ is defined in the 2D complex plane as a function of $\overrightarrow{r}$
\begin{equation}
\langle \, \overrightarrow{n} (\overrightarrow{r}) \cdot \overrightarrow{n} (\overrightarrow{r}^{\prime}) \, \rangle_{n} =
\langle \, e^{i \theta (\overrightarrow{r})} e^{ -i \theta (\overrightarrow{r}^{\prime})} \, \rangle_{n} \backsim
\left \vert \overrightarrow{r} - \overrightarrow{r}^{\prime}\right \vert^{-\eta}
\label{1}
\end{equation}
Here, the average is that defined in thermodynamic equilibrium and the orientational order decays algebraically \cite{9}, as is well known. The 2D wormlike chain embedded in the nematic is defined by $z (s) = x (s) + i \, y (s)$ where $s$ is a point on the contour from one end ($0 \leq s \leq N$). Because of the strong coupling, the chain is slaved to the nematic. The effective Hamiltonian $\cal{H}$ is a functional of $z$ and $\overrightarrow{n}$ and consists of the bending energy of the chain, the free energy of the fluctuating nematic and a term signifying the strong coupling of the chain to the nematic as discussed by Nelson {\em et al.} \cite{8,9} (for a qualitative treatment of the enslavement, see the Appendix). We have
\begin{equation}
\frac{dz}{ds} = e^{i \theta (s)}
\label{2}
\end{equation}
where the right-hand-side is conveniently regarded as a functional of $z (s)$. We therefore obtain
\begin{equation} \langle \, | z(s) - z(s^{\prime}) |^{2} \, \rangle_{nc} =
\int \limits_0^N \!\! ds \int \limits_0^N \!\! ds^{\prime} \; \langle \, | e^{i \theta  \left [z (s)\right ]} e^{ -i \theta \left [z (s^{\prime})\right ]} | \, \rangle_{nc}
\label{3}
\end{equation}
The index~$nc$ denotes that two averages~have been employed within a canonical ensemble, that is including a factor $\exp(- {\cal H} / k_{\rm B} T)$ where $k_{\rm B}$ is Boltzmann's constant and $T$ is the temperature. One average is a functional integration over fluctuations in the nematic ($n$), the other is a functional integration over chain configurations ($c$). Inserting eq (\ref{1}) into eq (\ref{3}), we end up with an integral equation. Upon setting $z (N) \backsim N^{\nu}$, we conclude that \cite{8,9}
\begin{equation}
\nu = \frac{2}{2 + \eta}
\label{4}
\end{equation}
It is remarkable that this expression is derived without having to use the probability $\exp(- {\cal H} / k_{\rm B} T)$ itself \cite{8,9}. The problem is whether an expression akin to eq (\ref{4}) is also valid for a material line in a 2D turbulent field.

A chain of length $N$ and mass $m_c$ immersed in a 2D Navier-Stokes fluid behaves like an unattached, one-dimensional flag acting on a fluid area of typical size $N^2$. One relevant dimensionless parameter $R_1 = m_{c} / N^{2} \rho_f$\ with $\rho_f$ the fluid density occurs in the theory of a singly attached flag flapping in an Euler fluid  \cite{10}.
A second parameter $R_2$ may be viewed as a non-dimensionalized bending energy. The bending energy $U_{b}$ is of order $B N / R_c^2$ where $R_c$ is the typical radius of curvature $R_c$ of the bent flag and $B$ is the bending force constant. The bending energy is at most $B / N$ so that the elastic energy density scales as $B / N^3$. We have to compare this with the fluid Reynolds stress $\rho _{f} U^{2}$ where $U$ is a typical velocity of the flag with respect to some background at the far field. We therefore have $R_2 = B / \rho_f U^2 N^3$. We wish to consider the limit where the back reaction of the flag on the fluid is negligible. Von Karman vortices arising at the two ends (when the flag is free) have little effect when $R_2 \gg R_1$  \cite{10,11}. On the other hand, an energy criterion $R_2 \ll 1$ has been introduced by de Gennes \cite{12} to ascertain when passive advection is valid in three dimensions. This criterion has also been applied to 2D turbulent flows  \cite{13}. A regime with both $R_2 \gg R_1$ and $R_2 \ll 1$ is easily realizable according to fig 3 of ref. \cite{10}.

Of course, a Navier-Stokes fluid is definitely not an Euler fluid even as the kinematic viscosity goes to zero \cite{14} but let us focus on the enstrophy cascade at very high Reynolds numbers. The turbulence is stationary and homogeneous. The rate of dissipation at scales smaller than the injection scale is $\chi  = d \langle \omega^2 \rangle_h / dt$ where $ \langle \text{} \rangle_h$ represents an average over an ensemble of realizations of the vorticity $\omega (\overrightarrow{r} ,t)$ \cite{5}. The inertial regime is here between $\lambda_k$ and the injection scale; it is scaleless. A material line swaying in the fluid has a viscous boundary layer of size $\lambda_k$ along its length. At a distance $l$ from this line, the largest eddy must be of order $l$ (at least if the radius of curvature $R_c$ is not too small). But the typical time scale of all the eddies including those in the turbulent boundary layer must be $\chi^{-1/3}$. If we suppose a power law for the material line holds again: $R \sim N^{\nu_h}$, passive advection implies full enslavement of the material line to the flow in the enstrophy cascade regime. I again stress that nowhere in the above analysis of the nematic problem leading to eq (\ref{4}) is explicit use made of a probability function within a canonical ensemble. Hence, one may apply the identical argumentation to a 2D turbulent stationary state with an unknown probability function pertaining to that state. Thus, we simply follow the above line of reasoning to write
\begin{equation}
\nu_h = \frac{2}{2 + \eta_h}
\label{5}
\end{equation}
where the exponent is defined in terms of the hydrodynamic velocity $\overrightarrow{v} (\overrightarrow{r}) \equiv v (\overrightarrow{r}) \overrightarrow{n}(\overrightarrow{r})$ which defines a ``polar'' director $\overrightarrow{n} (\overrightarrow{r})$
\begin{equation}
\langle \, \overrightarrow{n} (\overrightarrow{r}) \cdot \overrightarrow{n} (\overrightarrow{r}^{\prime}) \, \rangle_h \backsim
\left \vert \overrightarrow{r} - \overrightarrow{r}^{ \prime }\right \vert^{- \eta_h}
\label{6}
\end{equation}
The amplitude of the velocity vector is $v (\overrightarrow{r})$ and the index~$h$ denotes an average over an ensemble of stationary states.

The orientational correlation function given by eq (\ref{6}) appears to have never been computed; in principle, it may hold on general grounds in two dimensions since the Kraichnan regime is scaleless. In the experiments by Amarouchene and Kelly \cite{7}, the soap film flows on average in the $y$ direction under gravity. The fluctuation $h(y)$ of the injected material line consisting of pure water is measured in the direction perpendicular to the $y$ axis. Thus, it is expedient to focus on the correlation or structure functions $\langle | \delta h(r) |^n \rangle_h$ with $\delta h(r) \equiv h(y+r) - h(y)$. For $n \!=\! 2$, this function scales empirically as $r^{\xi_n}$ where the exponent $\xi_n$ is close to 2 at low rates of flow where the coherent vortices appearing in the 2D fluid are ordered. At higher rates of flow, the 2D film becomes turbulent and the coherent vortices are scattered throughout the turbulent background in a disordered manner. The exponent $\xi_2$ ultimately reaches a value of about 1.5 continuously until anomalies start to occur related to the integrity of the material line. In fig~1 of ref.~\cite{7}, the line seems to be attracted to coherent vortices here and there.

Although this issue was not investigated, it is probably safe to posit that the line fluctuations are isotropic implying $\xi_2 \equiv 2 \nu_h$. Accordingly, $\nu_h$ would range from unity at low rates of flow to about 3/4 at high rates. The relation between the exponent $\nu_h$ and $\eta_h$ given by eq (\ref{5}) can be tested purely empirically as suggested by Hamid Kellay (private communication). In ref.~\cite{7}, the exponent $\xi_2$ is a function of the Reynolds number but this may not mean much; the turbulence in the inertial Kraichnan regime could be fully developed whereas the coherent vortices and their distribution could well still depend on the viscosity of the soap film. Another potential problem in the scaling analysis is that the dimensionless coefficients $R_1$ and $R_2$ may need to be renormalized if a power law for the chain size $R(N)$ is posited. Nevertheless, the interaction between the material line and the Kraichnan fluid is strongly non-local. The renormalization may be surmised to be less than in the nematic case where the chain fluctuates in a heat bath and the stochastic forces on it are essentially point-like. 

\section*{Acknowledgment}

I would like to thank Hamid Kellay and Yacine Amarouchene for their hospitality and enlightening discussions. A query by Vincenzo Vitelli prompted me to write the Appendix.

\section*{Appendix}
\renewcommand{\theequation}{A\arabic{equation}}
\setcounter{equation}{0}

A qualitative analysis of a wormlike chain fluctuating within an inhomogeneous nematic in three dimensions (3D) was presented a long time ago \cite{15,16}. If the director varies slowly, the chain is basically aligned along it except for small undulations on the scale of the so-called deflection length $\lambda_3$ \cite{15}. Let us first focus, in 2D, on a small region where the director is effectively a constant vector. The chain of length $N$ is slaved to the director except for fluctuations given by a Gaussian
\begin{equation}
H \backsim \exp (- \tfrac{1}{2} \, \beta \, \phi^2)
\label{A1}
\end{equation}
where $\beta \!\gg\! 1$ and $\phi$ is the angle between a small segment of the chain and the director. In that case $\lambda_2 \!=\! P / \beta$ where $P$ is the 2D persistence length of the chain ($N \!\gg\! P$).

On the other hand, there is another coupling of the molecules of the pure nematic given by
\begin{equation}
G \backsim \exp (- \tfrac{1}{2} \, \alpha \, \psi^2)
\label{A2}
\end{equation}
where $\alpha \!\gg\! 1$ and $\psi$ is the angle between a molecule of length $a$ and the director. The scale $a$ is viewed as a short-distance cut-off. Within a continuum approximation we require $\lambda_2 \!\gg\! a$. Hence, the 2D suspension is bidisperse and consists of deflection segments of length $\lambda_2$ interacting with rods of length $a$. In the 3D case, one may show that $\alpha \!\gg\! \beta$ \cite{17} which is readily extended to the 2D case.

Next, we consider a much larger 2D space in which the director varies. Both splay and bend Frank elastic moduli are set equal to $K$ (which is scaled by $k_{\rm B} T$). At large wavelengths, the dimensionless free energy is equivalent to that of the XY model \cite{18} and is given by $\tfrac{1}{2} K \int \! d \vec{r} \; | \overrightarrow{\nabla} \theta |^2$. The orientational fluctuations are Gaussian so that eq (\ref{1}) becomes
\begin{equation}
\langle \, \overrightarrow{n} (\overrightarrow{r}) \cdot \overrightarrow{n} (\overrightarrow{r}^{\prime}) \, \rangle_{n} =
\exp \left( - \tfrac{1}{2} \langle \, (\Theta (\overrightarrow{r}) - \Theta (\overrightarrow{r}^{\prime}))^2 \, \rangle_{n} \right)
\label{A3}
\end{equation}
The second moment is computed via the equipartition theorem
\begin{equation}
\langle \, (\Theta (\overrightarrow{r}) - \Theta (\overrightarrow{r}^{\prime}))^2 \, \rangle_{n} \simeq
\frac{1}{\pi K} \, \ln \left( \frac{| \overrightarrow{r}- \overrightarrow{r}^{\prime} |}{a} \right)
\label{A4}
\end{equation}
Eqs (\ref{1}), (\ref{A3}) and (\ref{A4}) lead to
\begin{equation}
\eta = \frac{1}{2 \pi K}
\label{A5}
\end{equation}
However, the algebraic decay displayed in eq (\ref{1}) is only valid when the temperature is below a certain critical temperature as argued by Kosterlitz and Thouless \cite{18}. At higher temperatures the decay turns out to be exponential. The exponent $\eta$ has an upper bound equal to $1/4$ in the regime of algebraic decay.

I next derive scaling relations noting that $\alpha$ is irrelevant as shown above. At short enough distances ($\backsim \ell$), the chain must remain enclosed within a triangular region given by eq (\ref{A4}). We have 
\begin{equation}
\ell \simeq \exp \left( \frac{1}{\eta \beta} \right)
\label{A6}
\end{equation}
At large distances, the typical degree of inhomogeniety of the nematic $\partial \theta / \partial r$ is always smaller than $\eta \, \beta^{1/2} / \ell$ via eq (\ref{A4}). Therefore, the worm is slaved to the director except for undulations on a scale $\lambda_2$. Eq (\ref{A1}) then implies
\begin{equation}
P < \frac{\ell}{\eta}
\label{A7}
\end{equation}
This requirement is easily met at small $\eta$. It is noted that the orientational fluctuations ultimately lead to vortices in view of eq (\ref{A4}) \cite{18}.


\begin{thebibliography}{99}

\bibitem {1} A. M. Polyakov, Nucl. Phys. B \textbf{396}, 367 (1993).

\bibitem {2} A. M. Polyakov, Phys. Rev. \textbf{52}, 6183 (1995).

\bibitem {3} G. Falkovich, J. Phys. A \textbf{42}, 123001 (2009).

\bibitem {4} R. H. Kraichnan, Phys. Fluids \textbf{10}, 1417 (1967).

\bibitem {5} G. Batchelor, Phys. Fluids II supplement \textbf{12}, 233 (1969).

\bibitem {6} H. Kellay and W. I. Goldburg, Rep. Prog. Phys. \textbf{65}, 845 (2002).

\bibitem {7} Y. Amarouchene and H. Kellay, Phys. Rev. Lett. \textbf{95}, 054501 (2005).

\bibitem {8} R. D. Kamien, P. Le Doussel and D. R. Nelson, Phys. Rev. A \textbf{45}, 8727 (1992).

\bibitem {9} D. R. Nelson, Defects and geometry in condensed matter physics, Cambridge University Press, UK, 2002.

\bibitem {10} M. J. Shelley and J. Zhang, Annu. Rev. Fluid Med. \textbf{43}, 449 (2011).

\bibitem {11} S. Alben and M. J. Shelley, Phys. Rev. Lett. \textbf{100}, 074301 (2008).

\bibitem {12} P. G. de Gennes, Physica A \textbf{140}, 9 (1986).

\bibitem {13} T. Odijk, Physica A \textbf{298}, 140 (2001).

\bibitem {14} G. L. Eyink, Physica D \textbf{237}, 1956 (2008).

\bibitem {15} T. Odijk, Liquid Crystals \textbf{1}, 553 (1986).

\bibitem {16} A. Yu. Grosberg and A. V. Zhestkov, Vysokomol. Soed \textbf{28}, 86 (1986).

\bibitem {17} T. Odijk, Macromolecules \textbf{19}, 2313 (1986).

\bibitem {18} J. M. Kosterlitz and D. J. Thouless, J. Phys. C \textbf{6}, 1181 (1973).

\end{thebibliography}
\end{document}